\documentclass[twocolumn,noshowpacs,showkeys,preprintnumbers,amsmath,amssymb,prl]{revtex4}

\usepackage{upgreek}
\usepackage{color}

\usepackage{epsfig}
\usepackage{graphicx}
\usepackage{dcolumn}
\usepackage{bm}
\usepackage{bbm}
\usepackage{mathrsfs}
\usepackage{verbatim}
\usepackage{amsmath}
\usepackage{multirow}
\usepackage{feynmf}
\usepackage[all, knot]{xy}
\xyoption{arc}

\begin{document}

\title{Quantum computational path summation for relativistic quantum mechanics\\
and a time dilation relation for  a Dirac Hamiltonian generator on a qubit array}

\author{Jeffrey Yepez}

\address{
Department of Physics and Astronomy, University of  Hawai`i at Manoa, Honolulu, Hawai`i 96822
 }

\date{September 26, 2015, Revised August 4, 2017}

\begin{abstract}
Dirac particle dynamics is encoded as a unitary path summation rule and implemented on a qubit array, where the qubit array represents both spacetime and the fermions contained therein.  The unitary path summation  rule gives a quantum algorithm to model a many-body system of Dirac particles in a gauge field with Lorentz invariance down to the grid scale (Planck scale)---the lattice-based model neither suffers the Fermi-sign problem nor breaks Lorentz invariance.   Yet, for the Dirac Hamiltonian to generate the unitary evolution of the 4-spinor field at the Planck scale, there is time dilation between the shortest observable time near a single space point and that time measured at long-wavelength scales.  We find gravitational time dilation where the model space around each point (with an even number of qubits) is curved like  the space around a Schwarzschild black hole.
\end{abstract}


\maketitle

{\it Introduction.---}Quantum computation used for quantum simulation offers a new way to explore relativistic quantum mechanics and demonstrate
relativistic effects in the Dirac equation \cite{PhysRevLett.98.253005,1367-2630-15-7-073011,4734301720100107,PhysRevLett.106.060503}.   
Quantum lattice gas models have  been explored for relativistic quantum mechanics \cite{JSP.53.323,FPL.10.105,yepez-qip-05} and tested in one-body simulations in 1+1 dimensions \cite{meyer-JStatPhys96,meyer-pla96,meyer-PhysRevE97,PhysRevE.55.5261,meyer-jpamg98}.  
Quantum lattice gas  and  quantum Boltzmann models for relativistic quantum systems have  been investigated  for  quantum simulation applications \cite{succi-PhysRevE96,PhysRevC.84.034903,yepez_arXiv1307.3593_quant_ph,PhysRevLett.111.160602}. 

Presented here is a novel quantum information dynamics model of relativistic quantum mechanics expressed as a simple rule to calculate a Feynman path integral  in 3+1 dimensions. The probability amplitude, say $K$,  for a relativistic Dirac particle of mass $m$ to go from point $x_a^\mu=(c t_a, \bm{x}_a)$ to point  $x_b^\mu=(ct_b, \bm{x}_b)$  is  calculated by summing over all  paths that  connect the points   $a$ to $b$:
\begin{equation}
\label{Yepez_unitary_path_summation_high_energy}
K_{ab}^\text{\tiny HE}
 = 
\sum_{R\ge 0}\Phi_{ab}(R)
  (
\sqrt{
  1-
\epsilon^2}
 \, )^{\overline{R}}
  \left( -i\epsilon\right)^{R},
\end{equation}
where $\epsilon \equiv  mc\ell/{\hbar}$,  $R$ is the number of bends (in space) in a path,  $\overline{R}$ is the number of unbends, and  $\Phi_{ab}({R})$ is  the number of  paths with $R$ bends.  Once the start and end points are selected, the only free parameter is $m$, as  the reduced Planck constant $\hbar$,   Planck length $\ell$, Planck time $\tau$, and speed of  light $c=\ell/\tau$ are  constants.  A spacetime path can only be resolved up to the Planck scales,  so with integer $N=(t_b-t_a)/\tau$ time steps, the number of unbends is $\overline{R}\equiv N-R$.  The dimensionless parameter $\epsilon \equiv  mc\ell/{\hbar}\le 1$ is a small parameter  when the Planck length $\ell$ is  smaller than the reduced Compton wavelength $\hbar/mc$ of the Dirac particle. Yet, the path summation rule (\ref{Yepez_unitary_path_summation_high_energy})  models the particle physics all the way up to $\epsilon=1$,  called the highest energy (HE) limit.

Kernel (\ref{Yepez_unitary_path_summation_high_energy}) appears simple, yet it is rich in particle physics. It applies to antiparticles as well. By writing it in spin variables, (\ref{Yepez_unitary_path_summation_high_energy}) leads to an efficient quantum computing algorithm for modeling Dirac particles.  In this sense, it models relativistic quantum mechanics, including how the dynamical behavior of the fermionic 4-spinor field $\psi(x)$  is generated by the Dirac Hamiltonian. It  also explains the particle motion in a background 4-potential field $A^\mu(x)$. Even particle interactions mediated by exchange of quanta of the gauge field $A^\mu(x)$ is represented by  (\ref{Yepez_unitary_path_summation_high_energy}), and this quantum field theory application is given an another communication \cite{yepez_arXiv1612.09291_quant_ph}.  In this Letter,    (\ref{Yepez_unitary_path_summation_high_energy}) is evaluated on a 3+1 spacetime lattice for relativistic particle dynamics in a background 4-potential field.

Expanding in $\ell$,  $K_{ab}^\text{\tiny HE}= K_{ab}^{(1)}+K_{ab}^{(2)}+ \cdots$.  The lowest-order term 
 $K_{ab}^{(1)}=\sum_{R\ge 0}\Phi_{ab}(R) \left( -i\epsilon\right)^{R}$ is not unitary on a finite spacetime lattice with 4-volume $(T\tau) (L\ell)^3$. Yet, $K_{ab}^{(1)}$ becomes unitary in the continuum limit $L \rightarrow \infty$ of Minkowski space so long as the number of time steps is much smaller that the system size $N\lll L$
\begin{equation}
\label{yepez_feynman_path_summation}
K_{ab}^{(1)\text{Feynman}}
 = \lim_{L\rightarrow \infty}\sum_{R\ge 0}\Phi_{ab}(R) \left( -i\epsilon\right)^{R}.
\end{equation}
Analytically evaluating (\ref{yepez_feynman_path_summation}) on a 1+1 dimensional spacetime lattice is known as the Feynman chessboard problem \cite{feynman-65-1st-qlga,jacobson-jpamg84,jacobson-jpamg.v17.84}. The associated quantum lattice gas algorithm for kernel (\ref{yepez_feynman_path_summation}) on 3+1 dimensional spacetime lattice was previously found \cite{yepez-qip-05}, providing the first efficient quantum computational algorithm for  Dirac particle dynamics  \cite{yepez_arXiv1307.3593_quant_ph,PhysRevLett.111.160602}.     
 Yet, the unitarity of (\ref{Yepez_unitary_path_summation_high_energy})  requires no limiting process, and it becomes a relativistic path integral where Lorentz invariance is retained down to the Planck scale.  Therefore, it can be used to explore particle physics at the Planck scale. For example,  (\ref{Yepez_unitary_path_summation_high_energy}) predicts that Schwarzschild time-dilation occurs near this scale.  In short, this Letter presents a novel quantum information dynamics algorithm for  quantum simulation of relativistic quantum mechanics and presents the discovery of gravitational time-dilation that occurs at the highest-energy scales when the particle physics is given (\ref{Yepez_unitary_path_summation_high_energy}).

{\it Spin-chain encoded path.---}The particle physics  of a massless fermion of energy $E$ moving at the speed of light $c$  is described by the  $(
\sqrt{
  1-
\epsilon^2}
 \, )^{\overline{R}}$ term in (\ref{Yepez_unitary_path_summation_high_energy}); the chiral fermion's 4-momentum $p^\mu =  (p_{0}, \bm{p})$ is light like, $p_\mu p^\mu = 0$.
 The $\left( -i\epsilon\right)^{R}$ term in (\ref{Yepez_unitary_path_summation_high_energy}) encodes chiral symmetry breaking; the fermion's 4-momentum becomes $p^\mu=(E/c,mc \bm{u})$ and  $p_\mu p^\mu = mc^2$.   The proper velocity $\bm{u}$ can be parametrized by the spherical Euler angles $\theta$ and $\varphi$, viz. $\bm{u} = (\sin\theta\cos\varphi,\sin\theta\sin\varphi,\cos\theta)$.

 Casting (\ref{Yepez_unitary_path_summation_high_energy}) in spin variables provides a way to derive a quantum algorithm for relativistic quantum mechanics.  The spin space is a tensor product space over a discrete field of qubits. This discrete field is called a qubit array. So let  a qubit, say $|q_s\rangle$, encode the direction of $\bm{u}$. The path summation rule becomes a quantum information dynamics model.  The conventions of particle physics are inverted using a qubit to encode the spin state of a spin-1/2 particle:    the qubit is the fundamental object and a spin-1/2 particle is  a bit contained in the qubit.

All the translational degrees of freedom of a particle on the spacetime lattice are encoded in spin variables; the fermion's path from point $a$ to $b$ is specified by a chain of spin 4-vectors, where each 4-vector at a point is denoted
$s^\mu =(s_0,\bm{s}) = (s_0, s_1 , s_2 , s_3 )$.
The quartic  $s^\mu s_\mu = s_0^2 - s_1^2 - s_2^2 - s_3^2=0$ is light like \cite{yepez-qip-05} \footnote{The fermion's motion is encoded in spin variables $s_0=\pm 1$ (forward/backward  in time) and $(s_1,s_2, s_3)=(\pm 1,\pm 1, \pm 1)/\sqrt{3}$ (parallel/antiparallel  in space  to $\hat{\bm{x}}$, $\hat{\bm{y}}$, $\hat{\bm{z}}$, respectively) along the lattice directions of a body-centered cubic lattice.
}.  
Since qubits  contain the  fermionic bits, each bit can simultaneously occupy all points in the qubit array in quantum superposition representing how a fermion can occupy all points of space in superposition. 
Using the Pauli matrices
\begin{align}
\sigma_x = 
{\scriptsize
\begin{pmatrix}
  0    &   1 \\
 1     &  0
\end{pmatrix}
},
\qquad
\sigma_y = 
{\scriptsize
\begin{pmatrix}
  0    &   -i \\
 i     &  0
\end{pmatrix}
},
\qquad
\sigma_z = 
{\scriptsize
\begin{pmatrix}
  1    &   0 \\
 0     &  -1
\end{pmatrix}
}
\end{align}
 as the fundamental representation of the special unitary group SU(2) and spin-space 3-vector $\bm{\sigma} = (\sigma_x, \sigma_y, \sigma_z)$ and 4-vector $\sigma^\mu=(\bm{1}, \bm{\sigma})$.
 Using the spin-$1/2$  operator $\frac{\hbar}{2}\bm{\sigma}$, the proper velocity $\bm{u}$ may be cast in matrix form as a spin operator: $\bm{u}\mapsto  \hat{\bm{S}} \equiv \frac{\hbar}{2}\bm{\sigma} . {\bm{u}}$
\begin{equation}
\hat{\bm{S}}
=
{\frac{\hbar}{2}}
\begin{pmatrix}
u_{z} & u_{x}-iu_{y}\\
u_{x}+iu_{y} & -u_{z}
\end{pmatrix}
=
{\frac{\hbar}{2}}\left(\begin{array}{cc}
\cos\theta & \sin\theta e^{-i\varphi}\\
\sin\theta e^{i\varphi} & -\cos\theta\end{array}\right).
\end{equation}
The  spin eigenequation $\hat{\bm{S}} {|\pm\rangle}_{s}=\pm{\frac{\hbar}{2}}{|\pm\rangle}_{s}$ has  spin-1/2 eigenstates  
 \begin{subequations} 
 \label{eq:+-kets} 
 \begin{eqnarray}
{|+\rangle}_{s}&=&\; \;\;\cos{\frac{\theta}{2}} e^{-i{\frac{\varphi}{2}}}|0\rangle+\sin{\frac{\theta}{2}} e^{i{\frac{\varphi}{2}}}|1\rangle\label{eq:+ket}
\\
{|-\rangle}_{s}&=&-\sin{\frac{\theta}{2}} e^{-i{\frac{\varphi}{2}}}|0\rangle+\cos{\frac{\theta}{2}} e^{i{\frac{\varphi}{2}}}|1\rangle.
\label{eq:-ket}
\end{eqnarray}
 \end{subequations} 
Therefore, the fermion's 3-momentum $\bm{p} =\langle q_s| \frac{mc}{\hbar}\hat{\bm{S}}| q_s\rangle$ is encoded on the Bloch sphere   of a 2-level qubit as
 \begin{equation}
\label{Bloch_sphere_velocity_encoding}
  |q_s\rangle 
  \!\!
  =
e^{-i \frac{E \tau}{\hbar}}
\left[
  \cos\left(\frac{\vartheta}{2}\right)  |0\rangle
  +e^{i \varphi}  \sin\left( \frac{\vartheta}{2}\right)  |1\rangle
  \right]
  ,
\end{equation}
where the fermion's energy determines the overall phase. 
The fermion's wave 4-vector is   
   $k_n^\mu 
   \equiv     \frac{2\pi}{\ell}
 \left(
 \frac{n_t}{T}, 
 \frac{n_x}{L}, \frac{n_y}{L}, \frac{n_z}{L}
 \right)$, 
for integers $n_t$, $n_x$, $n_y$ and $n_z$ and for  system size $T$ in time and $L$ in space.  For a massive fermion, $k_n = \sqrt{k_{n\mu} k_n^\mu} \ne 0$. The fermion's 4-momentum is $p^\mu_n = \hbar k^\mu_n$.

The motion of a fermion/antifermion moving say from point $(ct_a, \bm{x}_a)$ to $(ct_b,\bm{x}_b)$ is constrained such that ${s}_0 {k}_0=\pm |k_0|$. The particle dynamics in (\ref{Yepez_unitary_path_summation_high_energy}) never changes a fermion to an antifermion (nor vice versa), so there is no loss of generality by restricting our attention to the dynamics of a positive-energy fermion. The treatment for negative-energy fermions is similar.

 Although a fermion's direction in time  does not change while it exists, its 3-momentum changes direction upon photon absorption/emission. The path summation (\ref{Yepez_unitary_path_summation_high_energy})  takes this process into account. At the $w$th time step, the fermion's local outgoing  4-momentum is
\begin{align}
\label{4_momentum_conservation_b}
p'^\mu_{n}(A)
=p^\mu_{n}
-
\frac{e }{ c} {A}^\mu_{nw},
\end{align}
which is the  incoming 4-momentum $p^\mu_{n}$ at the $w$th time step of the path in the spacetime lattice minus the photon's 4-momentum \footnote{
 At each point of its motion, the fermion can absorb a photon from the background Maxwell 4-potential field  $A^\mu(x)=(A_0(x), \bm{A}(x))$ while conserving 
energy $E'_{\bm{n}} =E_{\bm{n}} - e  {A}_{0\bm{n}}$ and 3-momentum ${\bm{p}}'_{\bm{n}} ={\bm{p}}_{\bm{n}} - {e }  {\bm{A}}_{\bm{n}}/c$. 
The 3-momentum  encoded in a qubit by (\ref{Bloch_sphere_velocity_encoding}) has  $|{\bm{p}}_{\bm{n}} | = |{\bm{p}}'_{\bm{n}} |$.
}.

During the $w$th time step ${s}_{0w}={s}_0$, and the fermion's momentum is ${p}^\mu_{n w} = (E/c{s}_{0}, mc {\bm{s}}_w) = (|p_{0n}| {s}_{0}, |\bm{p}_n| {\bm{s}}_w)$ in  spin variables. 
Moreover, a quanta of the background Maxwell field $A^\mu$ that interacts with this fermion during its $w$th time step is expressed  in spin variables  as
${A}^\mu_{nw}=  (A_{0nw} {s}_{0}, |\bm{A}_{nw}|  {\bm{s}}_w)$.
The fermion's interaction with a photon does not cause the fermion to reverse direction in time. The outgoing 4-momentum (\ref{4_momentum_conservation_b}) may be expressed in spin variables as
 \begin{align}
\label{4_vector_spin_operator_change}
   {s}'^\mu_w(A)
   &=
 {s}^\mu_w
 -
 \left(
\frac{e{A}_{0nw}}{|p_{0n}| c} 
,
\frac{e {\bm{A}}_{nw}}{|\bm{p}_{n}| c} 
 \right)
  .
\end{align}
Contracting (\ref{4_vector_spin_operator_change}) with ${p}_{n\mu}$ gives an identity needed later
 \begin{subequations} 
\label{spin_space_contraction_identity}
 \begin{align}
   {s}'^\mu_w(A) {p}_{n\mu}
 &=
 {s}^\mu_w  {p}_{n\mu}
 -
\frac{e{A}_{0nw} {p}_0}{|p_{0n}| c}   
+
\frac{e {\bm{A}}_{nw} \cdot {\bm{p}}_n}{|\bm{p}_{n}| c} 
\\
&=
 {s}^\mu_w   {p}_{n\mu}
 -
\frac{e{A}_{\mu nw}}{ c}   {s}^\mu_{w}
\\
&\stackrel{(\ref{4_momentum_conservation_b})}{=}
 {s}^\mu_w  {p}'_{n\mu}(A)
 .
\end{align}
 \end{subequations} 

The  displacement $x_b^\mu - x_a^\mu = (N, M_x, M_y, M_z)\ell$ of the  fermion on the spacetime lattice is specified by the positive integers $N$, $M_x$, $M_y$, $M_z$.  
 Let ${s}^\mu_0$  denote the  4-vector spin state of a Dirac particle at the initial point $x_a$ and ${s}^\mu_N$  the spin state at the end point $x_b$.  Moreover, let us consider the case when ${s}^\mu_0={s}^\mu_N$. 
The path summation (\ref{Yepez_unitary_path_summation_high_energy}) is equivalent to  a sum over a set of spin chains $\{s^\mu_0, \dots ,s^\mu_{N-1}\}$  with   4-magnetization $M^\mu= \ell\sum_{w=0}^{N-1} s^\mu_w$%
\begin{equation}
\label{yepez_feynman_path_summation-spin-form1}
K^\text{\tiny HE}_{ab}
=
\sum_{\{s^\mu_0, \dots ,s^\mu_{N-1}\}|M^\mu = C^\text{onst.}} 
\!\!
(
\sqrt{1-
\epsilon^2}
\,
)^{\overline{R}(A)}
\left(-{i\epsilon}\right)^{R(A)}.
\end{equation}
 Since
$\ell\sum_{w=0}^{N-1} s^\mu_w
= x^\mu_b-x^\mu_a$
, the constant magnetization spin chains are equivalent to fixed length paths  in spacetime.
 Also, with $s^\mu_0=s^\mu_{N}$, each spin chain is a closed loop (viz. periodic boundary conditions in spin variables). 

$R$ and $\overline{R}$ may be expressed in a contracted Lorentz product of $s^\mu$ 
at $w$ with $s_\mu$  at $w+1$ ($s'_\mu$ at $w$) as
\begin{equation}
\label{number_bends_and_unbends}
R(A)=
\frac{1}{2}
\sum_{w=0}^{N-1} s_w^\mu s_{\mu,w+1}
,
\;
\overline{R}(A) =
\sum_{w=0}^{N-1}
\left(
1- \frac{1}{2}
s_w^\mu s_{\mu,w+1}
\right)
.
\end{equation}
The  Kronecker delta is
 \begin{subequations} 
\label{kronecker_delta_for_fixed_magnetization_3d}
\begin{align}
\label{kronecker_delta_for_fixed_magnetization_3d_a}
\delta^{(4)}\left({{M}^\mu}, \sum_{w=0}^{N-1}{{s}}'^\mu_{w}\right)
&=
\frac{1}{T L^3}
\sum_{{n}}
e^{-i {x}_\mu  {k}^\mu_{n}}
e^{i 
{\ell}
\sum_w {s}'_{\mu w} {k}^\mu_{n}}
\\
\label{kronecker_delta_for_fixed_magnetization_3d_b}
\stackrel{(\ref{spin_space_contraction_identity})}{=}&
\frac{1}{TL^3}
\sum_{{n}}
e^{-i {x}_\mu  {k}^\mu_{{n}}}
e^{i 
{{\ell}}
\sum_w {s}_{\mu w}
\left( {k}^\mu_{{n}}
-
\frac{e {A}^\mu_{{n}w}}{\hbar c} 
\right)
}
\end{align}
 \end{subequations} 
in  3+1 dimensions \footnote{To emulate a fermion  in a  field $A =\sqrt{A_\mu A^\mu}\ne 0$,  one writes the  4-magnetization  
${M}^\mu=\sum_{w=0}^{N-1} {s}'^\mu_{w}$ in terms of the outgoing  spin   ${s}'^\mu_{w} =  {s}'^\mu_{w}(A)$  according to (\ref{4_vector_spin_operator_change}).  
Since the sum over a spin chain is a finite difference $\sum_{w=0}^{N-1} {s}'^\mu_w = \left(\frac{ct_b - ct_a}{{\ell}}, \frac{x_b - x_a}{{\ell}}, \frac{y_b - y_a}{{\ell}}, \frac{z_b - z_a}{{\ell}}\right)$, the Kronecker delta  is (\ref{kronecker_delta_for_fixed_magnetization_3d_a}).
} \footnote{ 
 As shorthand, the summation over wave vector modes is
$
\sum_{{n}}\equiv  \sum_{n_{t} = -T/2}^{(T/2)-1}  \sum_{n_{x} = -L/2}^{(L/2)-1} \sum_{n_{y} = -L/2}^{(L/2)-1} \sum_{n_{z} = -L/2}^{(L/2)-1}$. 
Cancellation of the path summation occurs outside of the light cone by the Kronecker delta (\ref{kronecker_delta_for_fixed_magnetization_3d}).  
}. 
Using 
$g \equiv -\frac{1}{2}\log\left(- {i\epsilon}\right)$
and 
$g' \equiv -\frac{1}{2}\log\left(\sqrt{1-\epsilon^2}\,\right)$, 
the kernel 
(\ref{yepez_feynman_path_summation-spin-form1}) 
becomes 
\begin{equation}
\label{yepez_feynman_path_summation_spin_form2}
\begin{split}
{K}^\text{\tiny HE}_{ab}
&\stackrel{(\ref{number_bends_and_unbends})}{=}
\sum_{\{s^\mu_0, \dots ,{s}^\mu_{N-1}\}}
\delta^{(4)}\left({M}^\mu,\sum_{w=0}^{N-1}{s}'^\mu_{w}(A)\right)
\\
&\times 
e^{\sum_{w=0}^{N-1}[- g {s}^\mu_{w} {s}_{\mu,w+1}-g'(2\bm{1}-{s}^\mu_{w} {s}_{\mu, w+1})]}
.
\end{split}
\end{equation}
 With ${s}^\mu_0={s}^\mu_N$ the spin chains are periodic, so $\sum_{w=0}^{N-1} {s}^\mu_w = \frac{1}{2}\sum_{w=0}^{N-1} ({s}^\mu_w + {s}^\mu_{w+1})$.  Inserting (\ref{kronecker_delta_for_fixed_magnetization_3d_b}) into (\ref{yepez_feynman_path_summation_spin_form2})
the kernel becomes a partition function of an ensemble of spin  chains 
\begin{align}
\label{yepez_feynman_path_summation_spin_form3}
{K}^\text{\tiny HE}_{ab}
&=
\frac{1}{T L^3}
\sum_{{n}}
e^{-i {x}_\mu  {k}^\mu_{\bm{n}}}
\sum_{\{s^\mu_0, \dots ,{s}^\mu_{N-1}\}}
 \prod_{w=0}^{N-1}
\\
\nonumber 
\times&
\underbrace{
e^{- i
\frac{\ell}{2}
({s}^\mu_w + {s}^\mu_{w+1})   {k}'_{\mu n}
- g {s}^\mu_{w} {s}_{\mu,w+1}-g'(2\bm{1}-{s}^\mu_{w} {s}_{\mu, w+1})
}
}
_{{{\cal U}}(n)}
.
\end{align}

{\it Quantum algorithm.}---At the $w$th time step in  (\ref{yepez_feynman_path_summation_spin_form3}), spacetime transfer operator  ${\cal U}(n)$ depends on the two 4-vectors $s^\mu_w$ and $s^\mu_{w+1}$, or $2^{4+4}$ combinations of spin variables.  Yet, by evaluating  ${\cal U}(n)$ with respect to the straight path connecting $x^\mu_a$ to $x^\mu_b$, the number of spin variables reduces to just $s_{0w}$, $s_{0,w+1}$, $\bm{s}_{w}\cdot \bm{k}_{\bm{n}}$ and $\bm{s}_{w+1}\cdot \bm{k}'_{\bm{n}}$.   In this frame, the $2^4$ combinations of  $\pm1$ variables allows  ${\cal U}(n)$ to be cast as a $4\times 4$ matrix and the $w$-product in (\ref{yepez_feynman_path_summation_spin_form3}) as matrix multiplication. 
In turn, the  transfer operator   is  ${{\cal U}}(n)= e^{-i\frac{(E-e A_{0w}) \tau}{\hbar}} 
{{\cal U}}(\bm{n})$, where 
\begin{eqnarray}
\label{natural_unbent_bent_chiral_basis}
{{\cal U}}(\bm{n})
=
 e^{- i
\frac{\ell}{2}
 ( {\bm{s}}_{w}
+ {\bm{s}}_{w+1}) \cdot  {\bm{{k}}}'_{\bm{n}}-g (1+{\bm{{s}}}_{w}\cdot{\bm{{s}}}_{w+1})-g' (1-{\bm{{s}}}_{w}\cdot{\bm{{s}}}_{w+1})}
\quad
\\
\nonumber
= 
\begin{cases}
e^{i
{{\ell}}
\bm{\sigma}\cdot {\bm{k}}_{\bm{n}}-2\mu\,\bm{1}}, &
\big(\bm{{s}}_{w}\cdot {\bm{k}}_{\bm{n}},\bm{{s}}_{w+1}\cdot {\bm{k}}'_{\bm{n}}\big)
=\big(-1,-1\big) 
\\
e^{-2\nu\,\bm{1}}, &  
\big(\bm{{s}}_{w}\cdot {\bm{k}}_{\bm{n}},\bm{{s}}_{w+1}\cdot {\bm{k}}'_{\bm{n}}\big)
=\big(-1,\;\;\;1\big) 
\\
e^{-2\nu\,\bm{1}}, &
\big(\bm{{s}}_{w}\cdot {\bm{k}}_{\bm{n}},\bm{{s}}_{w+1}\cdot {\bm{k}}'_{\bm{n}}\big)
=\big(\;\;\;\;\,1,-1\big) 
\\
e^{-i
{{\ell}}
\bm{\sigma}\cdot {\bm{k}}_{\bm{n}}-2\mu\,\bm{1}}, & 
\big(\bm{{s}}_{w}\cdot {\bm{k}}_{\bm{n}},\bm{{s}}_{w+1}\cdot {\bm{k}}'_{\bm{n}}\big)
=\big(\;\;\;\;\,1,\;\;\;1\big) .
\end{cases}
\end{eqnarray} 
So at the $w$th step for either unbent or bent path segment 
pairs $(\bm{{s}}_{w}\cdot {\bm{k}}_{\bm{n}},\bm{{s}}_{w+1}\cdot {\bm{k}}'_{\bm{n}})$, in matrix form (\ref{natural_unbent_bent_chiral_basis})  is
\begin{eqnarray}
{{\cal U}}(\bm{n})
\!\!\!
& =&
\nonumber
\!\!\!
\begin{pmatrix}
{{\cal U}}_{-1,-1} & {{\cal U}}_{-1,1} \cr
{{\cal U}}_{1,-1} & {{\cal U}}_{1,1}
\end{pmatrix}
=
\begin{pmatrix}
e^{i
{{\ell}}
 \bm{\sigma}\cdot {\bm{k}}'_{\bm{n}}-2\mu\,\bm{1}}
  &
e^{-2\nu\,\bm{1}}
  \cr
e^{-2\nu\,\bm{1}} 
& 
e^{-i
{{\ell}}
\bm{\sigma}\cdot{\bm{k}}'_{\bm{n}}-2\mu\,\bm{1}} 
\end{pmatrix}
\\
\label{unitary_transfer_matrix_3d_c}
=
&&
\hspace{-0.25in}
e^{i
{{\ell}}
 \sigma_z  \bm{\sigma}\cdot {\bm{k}}'_{\bm{n}} }
 \cdot
\begin{pmatrix}
\sqrt{1-\epsilon^2}\,\bm{1}
  &
-i\epsilon \,\bm{1}\cdot
e^{-i\ell
 \bm{\sigma}\cdot {\bm{k}}'_{\bm{n}}
}

  \cr
-i\epsilon\, \bm{1}\cdot
e^{ i\ell
 \bm{\sigma}\cdot {\bm{k}}'_{\bm{n}}
 }
& 
\sqrt{1-\epsilon^2}\,\bm{1}
\end{pmatrix},
\end{eqnarray}
using the convention  $ \sigma_z  \bm{\sigma}\equiv\sigma_z  \otimes \bm{\sigma}$ for tensor products.
This transfer operator constitutes an accurate quantum lattice gas algorithm,  the product of  stream  and  collide operators ${{\cal U}}(\bm{n})= {{\cal S}}(\bm{n})\cdot {{\cal C}}(\bm{n})$.  Since  $ {{\cal S}}(\bm{n})\equiv e^{i
{{\ell}}
 \sigma_z  \bm{\sigma}\cdot {\bm{k}}'_{\bm{n}} }$  is manifestly unitary, one can show the unitarity of ${{\cal U}}(\bm{n})$ by showing the unitarity of the collide operator
 \begin{subequations}
 \begin{align}
 {{\cal C}}(\bm{n}) 
 &\stackrel{(\ref{unitary_transfer_matrix_3d_c})}{=}
\sqrt{1-\epsilon^2}\,\bm{1}_4
-
i\epsilon
\sigma_x\bm{1}
\cdot
e^{ i\ell
\sigma_z  \bm{\sigma}\cdot {\bm{k}}'_{\bm{n}} 
}
\label{3D_path_collide_operator_c}
\\
&=
\exp\left[{- i \cos^{-1}(\sqrt{1-\epsilon^2})
\sigma_x\bm{1}
\cdot
e^{ i\ell
\sigma_z  \bm{\sigma}\cdot {\bm{k}}'_{\bm{n}} 
 }
 }
 \right],
 \label{3D_path_collide_operator_d}
\end{align}
\end{subequations}
 using  the identity $\sin(\cos^{-1}\sqrt{1-\epsilon^2})= \epsilon$ and realizing  $\sigma_x\bm{1}
\cdot
e^{ i\ell
\sigma_z  \bm{\sigma}\cdot {\bm{k}}'_{\bm{n}} 
 }
$ is an idempotent  operator. Multiplying (\ref{3D_path_collide_operator_c}) by ${{\cal S}}(\bm{n})$ 
  gives an accurate quantum algorithm  useful for simulating Dirac particle dynamics  
\begin{align}
\label{U_n_mass_form}
{{\cal U}}(\bm{n}) 
&=
\sqrt{1-\epsilon^2}
\,
{{\cal S}}(\bm{n})
-
i\epsilon
\sigma_x\bm{1}.
\end{align}
 This operator captures the quantum information dynamics intrinsic to relativistic quantum mechanics. 

{\it Dirac Hamiltonian.}---One is free to write the relativistic energy relation $E'^2 = (\hbar c\bm{k}'_{\bm{n}})^2 + (m c^2)^2$ (and using $\epsilon = mc^2 \tau/\hbar$) in spin space  as
\begin{equation}
(1-\epsilon^2)\bm{1}_4 =\left( 1-\left(\frac{E'\tau}{\hbar}\right)^2\right)\bm{1}_4 + \left( c\tau \sigma_z  \bm{\sigma}\cdot {\bm{k}}'_{\bm{n}}\right)^2
.
\end{equation}
Its square root with respect to complex conjugation is
\begin{equation}
\label{Yepez_relativistic_energy_relation_square_root}
\sqrt{1-\epsilon^2}\,
{{\cal S}}(\bm{n})
=
\sqrt{1-\left(\frac{E'\tau}{\hbar}\right)^2}
+ i
\ell \sigma_z \bm{\sigma}\cdot {\bm{k}}'_{\bm{n}}.
\end{equation}
Using ${\bm{p}}'_{\bm{n}}=\hbar {\bm{k}}'_{\bm{n}}$ and inserting (\ref{Yepez_relativistic_energy_relation_square_root}) into (\ref{U_n_mass_form}), the  spacetime transfer operator is   unitary 
 \begin{subequations}
\begin{align}
{{\cal U}}(n)
&=
e^{-i\ell  k'_0} 
\left[
\sqrt{1-\left(\frac{E'\tau}{\hbar}\right)^2}
+ i
\ell \sigma_z \bm{\sigma}\cdot {\bm{k}}'_{\bm{n}}
-
i\epsilon
\sigma_x\bm{1}
\right]
\\
=&
e^{-i \frac{(E-eA_0)\tau}{\hbar}}
\exp\left[
-i \cos^{-1}\left(\sqrt{1-\text{\tiny $\left(\frac{E'\tau}{\hbar}\right)$}^2}
\right)\frac{
{{h}}_\text{D}(\bm{n})
}{E'} \right]
.
 \label{fundamental_unitary_evolution_operator}
\end{align}
\end{subequations}
This demonstrates  that the unitary evolution from (\ref{Yepez_unitary_path_summation_high_energy})
is generated  without  approximation by   the Dirac Hamiltonian 
${{h}}_\text{D}(\bm{n})
=
-\sigma_z \bm{\sigma}\cdot ({\bm{p}}_{\bm{n}}c - e \bm{A})
+
\sigma_x\bm{1} \, mc^2 $.

{\it Time dilation.---}The transfer operator (\ref{fundamental_unitary_evolution_operator}) is
\begin{equation}
\label{QLG_QFT_Algorithm_HE}
{{\cal U}}(\bm{n}) =
 e^{-i \frac{(E-eA_0)\tau}{\hbar}}
 e^{- \frac{i}{\hbar}\zeta \tau {{h}}_\text{D}(\bm{n})},
\end{equation}
 where  $\zeta=\cos^{-1}\left(\sqrt{1-E'^2\tau^2/\hbar^2}
\right)/E'$ is a dimensionless time scale factor, which can be written as
\begin{equation}
\label{M_tau_dilation_equation}
\frac{E' \tau}{\hbar }  =  
\sin\frac{ E'  \zeta\tau}{\hbar}.
\end{equation}
At the grid-scale $E=\hbar/\tau$,  (\ref{M_tau_dilation_equation}) reduces to $1=\sin\zeta$, so here $\zeta ={\pi}/{2}$. 
Yet, at QFT scales $E\lll\hbar/\tau$, (\ref{M_tau_dilation_equation}) implies $\zeta \mapsto 1$.   So the scale range is $1\le \zeta\le {\pi}/{2}$. This implies that  the smallest observable intervals are the radial distance $r=\zeta\ell$ and elapsed time $t_r =  r/c = \zeta \tau$.

At immediate scales between the Planck scale $E=\hbar/\tau$ and  the QFT scale $E\lll\hbar/\tau$, and in the case where there are so many fermions in the system that the fields are well described in the mean-field limit, a low-energy expansion (\ref{M_tau_dilation_equation}) takes the form
\begin{equation}
\label{M_tau_dilation_equation_expanded_form2}
\frac{1 }{\zeta} =  
1- \frac{\zeta^2 \tau^2}{3\hbar^2 v^2/(p'^2 c^4)}+\cdots,
\end{equation}
where relativistic relation $v E' = p' c^2$  is used and $v<c$ is the fermion's velocity in the medium \footnote{
In the small-$E$ expansion,  rescaling $\tau \rightarrow \sqrt{2}\tau$ is applied in the equation for $\zeta$.  So  (\ref{M_tau_dilation_equation}) becomes
$\sqrt{2} \tau  =  
\frac{\hbar}{E}\sin\frac{\zeta E  \sqrt{2} \tau}{\hbar} = \sqrt{2}\zeta \tau\left(1- \frac{\zeta^2E^2 \tau^2}{3\hbar^2}+\cdots\right)$, 
and then both sides are divided by $\sqrt{2}\tau$.
}.   
In a many-body Fermi gas (e.g. Fermi condensate), the fermion's velocity (``acoustic" velocity) is less than the speed of light by a factor of $1/\sqrt{3}$ due to fermion-fermion  interactions.   
Upon rearranging,  taking the square root and setting $v = c/\sqrt{3}$, (\ref{M_tau_dilation_equation}) becomes a time dilation equation
\begin{equation}
\label{Yepez_Schwarzschild_dilation_equation}
\frac{\zeta \tau }{\hbar/(p'c)}=\sqrt{1-\frac{1}{\zeta}}
\quad
\longrightarrow
\quad
\frac{t_r}{t}  + \cdots =  
\sqrt{1-\frac{r_s}{r}},
\qquad
\end{equation}
where  
  the  Schwarzchild radius is $r_s = \ell$ and  in the long-wavelength limit the time is $t={\hbar}/{(p'c)}$.

{\it Path integral.---}At QFT scales ($\zeta\cong1$), (\ref{yepez_feynman_path_summation_spin_form3})  becomes
\begin{equation}
\label{yepez_feynman_path_summation_spin_form4}
{K}^\text{\tiny HE}_{ab}
\stackrel{(\ref{fundamental_unitary_evolution_operator})}{=}
\frac{1}{T L^3}
\sum_{n}
e^{- i x_\mu k^\mu_{n}}
\sum_{\text{paths}}
e^{-\frac{i}{\hbar}  \sum_{w=0}^{N-1} \delta t \,(E-eA_0- {h}_\text{\tiny D})}
.
\end{equation}
First, in the flat-space $L\rightarrow \infty$ continuum limit (with $\ell \sim dx\sim dy\sim dz$ and ${2\pi}/(L\ell)\rightarrow dk = {dp}/{\hbar}$),
the summations over $n=(n_t, n_x, n_y, n_z)$ and paths connecting $x^\mu_a$ to $x^\mu_b$ in 
(\ref{yepez_feynman_path_summation_spin_form4}) map  to momentum-space and path integrals, respectively,
\begin{align}
\sum_{n}\frac{1}{(2\pi)^4} \left(\frac{2\pi}{L\ell}\right)^4 
\sum_{\text{paths}}
\ell^3
&\mapsto
\int  \frac{d^4p}{(2\pi \hbar)^4} \int_a^b {\cal D}\{ \bm{x}(t) \}.
\label{path_integral_map}
\end{align}
Second, regarding  stream operator ${{\cal S}}(\bm{n}) 
=e^{ i\ell \sigma_z
 \bm{\sigma}\cdot {\bm{p}}_{\bm{n}} /\hbar
}$, 
where ${\bm{p}}_{\bm{n}}  = 2\pi\hbar \bm{n}/(L\ell)$, 
if  the  continuum limit $L\rightarrow \infty$ in a  many-fermion system is invoked, then
one replaces qubit-array operators ${\bm{p}}_{\bm{n}}$  and ${E}_{\bm{n}}$ 
  by operators
  acting on a  Dirac spinor field $\psi(x)$ in Minkowski spacetime
$\bm{{p}}_{\bm{n}} \mapsto -i\hbar\nabla$ 
and
${E}_{\bm{n}}
 \mapsto i\hbar\partial_t$. 
The  transfer operator maps to
\begin{equation}
\label{transfer_operator_highest_energy_quantum_algorithm_form}
{{\cal U}}({n})
\stackrel{(\ref{U_n_mass_form})}{\mapsto}
 e^{-i\frac{(E-e A_0) \tau}{\hbar}} 
\left[
\sqrt{1-\epsilon^2}\,
e^{ \ell  \sigma_z
 \bm{\sigma}\cdot
 \left(
 \nabla+ i \frac{e \bm{A}(x)}{\hbar c}
 \right)}
-
i\epsilon
\sigma_x\bm{1}
\right].
\end{equation}
$A_0(x)$ causes an overall phase rotation and 
 $\bm{A}(x)$  causes a phase rotation during streaming by ${{\cal S}}(\bm{n})$.  
In the 
 continuum  $L\rightarrow \infty$ limit of Minkowski space, the kernel (\ref{yepez_feynman_path_summation_spin_form4})     becomes a path integral
\begin{align}
\label{feynman_path_integral_in_p_space}
{K}^\text{\tiny HE}_{ab}
&\stackrel{(\ref{path_integral_map})}{\mapsto}
\int\frac{dp^4}{(2\pi\hbar)^4}
\int_a^b {\cal D}\{ x\} \,
e^{-i\frac{{x}_\mu {p}^\mu}{\hbar}}
 e^{-\frac{i}{\hbar} \int dt\,  \mathbb{L}(x)
 } 
,
\end{align}
%
%
where the Lagrangian operator $\mathbb{L}$ is given by 
\begin{equation}
\label{Lagrangian_operator_Legendre_form}
\mathbb{L}(x)
\equiv
{E}'_{n}
 -  {{h}}_\text{D}
 =
 \left[(\gamma^0 
{E}'_{\bm{n}}- (c\bm{\gamma} \cdot \bm{{p}}'_{\bm{n}}  +   mc^2)\right],
\end{equation}
and where  
the  chiral representation of the Dirac matrices \cite{Peskin_Schroeder_2004}  $\gamma_0\equiv \sigma_x \bm{1}_2 $ and $\bm{\gamma}\equiv i \sigma_y{\bm{\sigma}}$ is recovered.
    Since   
$\bm{{p}}'_{\bm{n}}
\mapsto-i\hbar \nabla - e \bm{A}/c$ 
and 
$ 
{E}'_{n}
\mapsto i \hbar \partial_t - e A_0= i \hbar  c\partial_0  + e A_0$, the Lagrangian operator (\ref{Lagrangian_operator_Legendre_form}) maps to
\begin{eqnarray}
\nonumber
\mathbb{L}
& \mapsto & \gamma_0  \left[ \gamma_0 \left(i \hbar  c\,\partial_0  -e A_0\right) 
-
c \bm{\gamma} \cdot\left(-i\hbar \nabla- e \bm{A}/c \right)-   mc^2\right]
\\
 \label{Lagrangian_operator}
& = & \gamma_0  \left[i\hbar c \,\gamma^\mu \left(\partial_\mu  + i \frac{eA_\mu}{\hbar c}\right) -   mc^2\right],
\end{eqnarray}
where $\gamma^\mu = (\gamma^0, \bm{\gamma})$, $\partial_\mu = (\partial_0, \nabla)$ and $A_\mu = (A_0, -\bm{A})$, reverting back to the  convention that matrix multiplication of two Dirac matrices does not require an infixed centered dot symbol.  
The Lagrangian density is defined in terms of the Lagrangian operator as
${\cal L}
\equiv
\psi^\dagger {\hat{L}} 
\psi$. 
Inserting (\ref{Lagrangian_operator}) into $\cal L$,
one obtains the Lagrangian density
${\cal L}_\text{D}
\equiv
\psi^\dagger {\hat{L}} 
\psi
\stackrel{(\ref{Lagrangian_operator})}{\mapsto}
\overline{\psi} \left(i\hbar c \,\gamma_\mu D^\mu -   mc^2\right) \psi$ as the effective quantum field theory for the quantum lattice gas. 
This is the covariant Lagrangian density for a Dirac 4-spinor field in the chiral representation,  where chiral symmetry is broken by the mass term, and where $D^\mu \equiv \partial^\mu + i e A^\mu(x)/(\hbar c)$ is the generalized 4-derivative with a background 4-potential $A^\mu(x)$. The matrix element of (\ref{yepez_feynman_path_summation_spin_form4})  takes the form of a Feynman path integral
$\langle \hat{K}_{ab}
\rangle
\mapsto 
\int_a^b{\cal D}\{ x\} \,
\int\frac{dp^4}{(2\pi\hbar)^4}
e^{\frac{i}{\hbar} \int d^4x \,{\cal L}_\text{D}}$. 

{\it Acknowledgements.---}This work was supported by the grant  ``Quantum Computational Mathematics for Efficient Computational Physics"  from the Air Force Office of Scientific Research. 


\begin{thebibliography}{21}
\expandafter\ifx\csname natexlab\endcsname\relax\def\natexlab#1{#1}\fi
\expandafter\ifx\csname bibnamefont\endcsname\relax
  \def\bibnamefont#1{#1}\fi
\expandafter\ifx\csname bibfnamefont\endcsname\relax
  \def\bibfnamefont#1{#1}\fi
\expandafter\ifx\csname citenamefont\endcsname\relax
  \def\citenamefont#1{#1}\fi
\expandafter\ifx\csname url\endcsname\relax
  \def\url#1{\texttt{#1}}\fi
\expandafter\ifx\csname urlprefix\endcsname\relax\def\urlprefix{URL }\fi
\providecommand{\bibinfo}[2]{#2}
\providecommand{\eprint}[2][]{\url{#2}}

\bibitem[{\citenamefont{Lamata et~al.}(2007)\citenamefont{Lamata, Le\'on,
  Sch\"atz, and Solano}}]{PhysRevLett.98.253005}
\bibinfo{author}{\bibfnamefont{L.}~\bibnamefont{Lamata}},
  \bibinfo{author}{\bibfnamefont{J.}~\bibnamefont{Le\'on}},
  \bibinfo{author}{\bibfnamefont{T.}~\bibnamefont{Sch\"atz}}, \bibnamefont{and}
  \bibinfo{author}{\bibfnamefont{E.}~\bibnamefont{Solano}},
  \bibinfo{journal}{Phys. Rev. Lett.} \textbf{\bibinfo{volume}{98}},
  \bibinfo{pages}{253005} (\bibinfo{year}{2007}),
  \urlprefix\url{http://link.aps.org/doi/10.1103/PhysRevLett.98.253005}.

\bibitem[{\citenamefont{LeBlanc et~al.}(2013)\citenamefont{LeBlanc, Beeler,
  Jim\'enez-Garc\'ia, Perry, Sugawa, Williams, and
  Spielman}}]{1367-2630-15-7-073011}
\bibinfo{author}{\bibfnamefont{L.~J.} \bibnamefont{LeBlanc}},
  \bibinfo{author}{\bibfnamefont{M.~C.} \bibnamefont{Beeler}},
  \bibinfo{author}{\bibfnamefont{K.}~\bibnamefont{Jim\'enez-Garc\'ia}},
  \bibinfo{author}{\bibfnamefont{A.~R.} \bibnamefont{Perry}},
  \bibinfo{author}{\bibfnamefont{S.}~\bibnamefont{Sugawa}},
  \bibinfo{author}{\bibfnamefont{R.~A.} \bibnamefont{Williams}},
  \bibnamefont{and} \bibinfo{author}{\bibfnamefont{I.~B.}
  \bibnamefont{Spielman}}, \bibinfo{journal}{New Journal of Physics}
  \textbf{\bibinfo{volume}{15}}, \bibinfo{pages}{073011}
  (\bibinfo{year}{2013}),
  \urlprefix\url{http://stacks.iop.org/1367-2630/15/i=7/a=073011}.

\bibitem[{\citenamefont{Gerritsma et~al.}(2010)\citenamefont{Gerritsma,
  Kirchmair, ZŠhringer, Solano, Blatt, and Roos}}]{4734301720100107}
\bibinfo{author}{\bibfnamefont{R.}~\bibnamefont{Gerritsma}},
  \bibinfo{author}{\bibfnamefont{G.}~\bibnamefont{Kirchmair}},
  \bibinfo{author}{\bibfnamefont{F.}~\bibnamefont{ZŠhringer}},
  \bibinfo{author}{\bibfnamefont{E.}~\bibnamefont{Solano}},
  \bibinfo{author}{\bibfnamefont{R.}~\bibnamefont{Blatt}}, \bibnamefont{and}
  \bibinfo{author}{\bibfnamefont{C.~F.} \bibnamefont{Roos}},
  \bibinfo{journal}{Nature} \textbf{\bibinfo{volume}{463}}, \bibinfo{pages}{68
  } (\bibinfo{year}{2010}), ISSN \bibinfo{issn}{00280836},
  \urlprefix\url{http://eres.library.manoa.hawaii.edu/login?url=http://search.ebscohost.com/login.aspx?direct=true&db=aph&AN=47343017&site=ehost-live}.

\bibitem[{\citenamefont{Gerritsma et~al.}(2011)\citenamefont{Gerritsma, Lanyon,
  Kirchmair, Z\"ahringer, Hempel, Casanova, Garc\'ia-Ripoll, Solano, Blatt, and
  Roos}}]{PhysRevLett.106.060503}
\bibinfo{author}{\bibfnamefont{R.}~\bibnamefont{Gerritsma}},
  \bibinfo{author}{\bibfnamefont{B.~P.} \bibnamefont{Lanyon}},
  \bibinfo{author}{\bibfnamefont{G.}~\bibnamefont{Kirchmair}},
  \bibinfo{author}{\bibfnamefont{F.}~\bibnamefont{Z\"ahringer}},
  \bibinfo{author}{\bibfnamefont{C.}~\bibnamefont{Hempel}},
  \bibinfo{author}{\bibfnamefont{J.}~\bibnamefont{Casanova}},
  \bibinfo{author}{\bibfnamefont{J.~J.} \bibnamefont{Garc\'ia-Ripoll}},
  \bibinfo{author}{\bibfnamefont{E.}~\bibnamefont{Solano}},
  \bibinfo{author}{\bibfnamefont{R.}~\bibnamefont{Blatt}}, \bibnamefont{and}
  \bibinfo{author}{\bibfnamefont{C.~F.} \bibnamefont{Roos}},
  \bibinfo{journal}{Phys. Rev. Lett.} \textbf{\bibinfo{volume}{106}},
  \bibinfo{pages}{060503} (\bibinfo{year}{2011}),
  \urlprefix\url{http://link.aps.org/doi/10.1103/PhysRevLett.106.060503}.

\bibitem[{\citenamefont{'t~Hooft}(1988)}]{JSP.53.323}
\bibinfo{author}{\bibfnamefont{G.}~\bibnamefont{'t~Hooft}},
  \bibinfo{journal}{J. Stat. Phys.} \textbf{\bibinfo{volume}{53}},
  \bibinfo{pages}{323} (\bibinfo{year}{1988}).

\bibitem[{\citenamefont{'t~Hooft}(1997)}]{FPL.10.105}
\bibinfo{author}{\bibfnamefont{G.}~\bibnamefont{'t~Hooft}},
  \bibinfo{journal}{Found. Phys. Lett.} \textbf{\bibinfo{volume}{10}},
  \bibinfo{pages}{105} (\bibinfo{year}{1997}).

\bibitem[{\citenamefont{Yepez}(2005)}]{yepez-qip-05}
\bibinfo{author}{\bibfnamefont{J.}~\bibnamefont{Yepez}},
  \bibinfo{journal}{Quantum Information Processing}
  \textbf{\bibinfo{volume}{4}}, \bibinfo{pages}{471} (\bibinfo{year}{2005}).

\bibitem[{\citenamefont{Meyer}(1996{\natexlab{a}})}]{meyer-JStatPhys96}
\bibinfo{author}{\bibfnamefont{D.~A.} \bibnamefont{Meyer}},
  \bibinfo{journal}{Journal of Statistical Physics}
  \textbf{\bibinfo{volume}{85}}, \bibinfo{pages}{551}
  (\bibinfo{year}{1996}{\natexlab{a}}).

\bibitem[{\citenamefont{Meyer}(1996{\natexlab{b}})}]{meyer-pla96}
\bibinfo{author}{\bibfnamefont{D.~A.} \bibnamefont{Meyer}},
  \bibinfo{journal}{Physics Letters A} \textbf{\bibinfo{volume}{223}},
  \bibinfo{pages}{337} (\bibinfo{year}{1996}{\natexlab{b}}).

\bibitem[{\citenamefont{Meyer}(1997{\natexlab{a}})}]{meyer-PhysRevE97}
\bibinfo{author}{\bibfnamefont{D.~A.} \bibnamefont{Meyer}},
  \bibinfo{journal}{Physical Review E} \textbf{\bibinfo{volume}{55}},
  \bibinfo{pages}{5261} (\bibinfo{year}{1997}{\natexlab{a}}).

\bibitem[{\citenamefont{Meyer}(1997{\natexlab{b}})}]{PhysRevE.55.5261}
\bibinfo{author}{\bibfnamefont{D.~A.} \bibnamefont{Meyer}},
  \bibinfo{journal}{Phys. Rev. E} \textbf{\bibinfo{volume}{55}},
  \bibinfo{pages}{5261} (\bibinfo{year}{1997}{\natexlab{b}}).

\bibitem[{\citenamefont{Meyer}(1998)}]{meyer-jpamg98}
\bibinfo{author}{\bibfnamefont{D.~A.} \bibnamefont{Meyer}},
  \bibinfo{journal}{Journal of Physics A: Mathematical and General}
  \textbf{\bibinfo{volume}{31}}, \bibinfo{pages}{2321} (\bibinfo{year}{1998}).

\bibitem[{\citenamefont{Succi}(1996)}]{succi-PhysRevE96}
\bibinfo{author}{\bibfnamefont{S.}~\bibnamefont{Succi}},
  \bibinfo{journal}{Physical Review E} \textbf{\bibinfo{volume}{53}},
  \bibinfo{pages}{1969} (\bibinfo{year}{1996}).

\bibitem[{\citenamefont{Romatschke et~al.}(2011)\citenamefont{Romatschke,
  Mendoza, and Succi}}]{PhysRevC.84.034903}
\bibinfo{author}{\bibfnamefont{P.}~\bibnamefont{Romatschke}},
  \bibinfo{author}{\bibfnamefont{M.}~\bibnamefont{Mendoza}}, \bibnamefont{and}
  \bibinfo{author}{\bibfnamefont{S.}~\bibnamefont{Succi}},
  \bibinfo{journal}{Phys. Rev. C} \textbf{\bibinfo{volume}{84}},
  \bibinfo{pages}{034903} (\bibinfo{year}{2011}),
  \urlprefix\url{http://link.aps.org/doi/10.1103/PhysRevC.84.034903}.

\bibitem[{\citenamefont{Yepez}(2013)}]{yepez_arXiv1307.3593_quant_ph}
\bibinfo{author}{\bibfnamefont{J.}~\bibnamefont{Yepez}},
  \bibinfo{journal}{arXiv:1307.3593 [quant-ph]}  (\bibinfo{year}{2013}).

\bibitem[{\citenamefont{Fillion-Gourdeau
  et~al.}(2013)\citenamefont{Fillion-Gourdeau, Herrmann, Mendoza, Palpacelli,
  and Succi}}]{PhysRevLett.111.160602}
\bibinfo{author}{\bibfnamefont{F.}~\bibnamefont{Fillion-Gourdeau}},
  \bibinfo{author}{\bibfnamefont{H.~J.} \bibnamefont{Herrmann}},
  \bibinfo{author}{\bibfnamefont{M.}~\bibnamefont{Mendoza}},
  \bibinfo{author}{\bibfnamefont{S.}~\bibnamefont{Palpacelli}},
  \bibnamefont{and} \bibinfo{author}{\bibfnamefont{S.}~\bibnamefont{Succi}},
  \bibinfo{journal}{Phys. Rev. Lett.} \textbf{\bibinfo{volume}{111}},
  \bibinfo{pages}{160602} (\bibinfo{year}{2013}),
  \urlprefix\url{http://link.aps.org/doi/10.1103/PhysRevLett.111.160602}.

\bibitem[{\citenamefont{Yepez}(2016)}]{yepez_arXiv1612.09291_quant_ph}
\bibinfo{author}{\bibfnamefont{J.}~\bibnamefont{Yepez}},
  \bibinfo{journal}{arXiv:1612.09291 [quant-ph]}  (\bibinfo{year}{2016}).

\bibitem[{\citenamefont{Feynman and Hibbs}(1965)}]{feynman-65-1st-qlga}
\bibinfo{author}{\bibfnamefont{R.~P.} \bibnamefont{Feynman}} \bibnamefont{and}
  \bibinfo{author}{\bibfnamefont{A.}~\bibnamefont{Hibbs}},
  \emph{\bibinfo{title}{Quantum Mechanics and Path Integrals}}
  (\bibinfo{publisher}{McGraw-Hill}, \bibinfo{year}{1965}),
  \bibinfo{note}{problem 2-6 on page 34.}

\bibitem[{\citenamefont{Jacobson and Schulman}(1984)}]{jacobson-jpamg84}
\bibinfo{author}{\bibfnamefont{T.}~\bibnamefont{Jacobson}} \bibnamefont{and}
  \bibinfo{author}{\bibfnamefont{L.}~\bibnamefont{Schulman}},
  \bibinfo{journal}{Journal of Physics A: Math. Gen.}
  \textbf{\bibinfo{volume}{17}}, \bibinfo{pages}{375} (\bibinfo{year}{1984}).

\bibitem[{\citenamefont{Jacobson}(1984)}]{jacobson-jpamg.v17.84}
\bibinfo{author}{\bibfnamefont{T.}~\bibnamefont{Jacobson}},
  \bibinfo{journal}{Journal of Physics A: Math. Gen.}
  \textbf{\bibinfo{volume}{17}}, \bibinfo{pages}{2433} (\bibinfo{year}{1984}).

\bibitem[{\citenamefont{Peskin and Schroeder}(1995)}]{Peskin_Schroeder_2004}
\bibinfo{author}{\bibfnamefont{M.~E.} \bibnamefont{Peskin}} \bibnamefont{and}
  \bibinfo{author}{\bibfnamefont{D.~V.} \bibnamefont{Schroeder}},
  \emph{\bibinfo{title}{An Introduction to Quantum Field Theory}}
  (\bibinfo{publisher}{Westview Press of the Perseus Books Group},
  \bibinfo{address}{New York}, \bibinfo{year}{1995}), \bibinfo{edition}{1st}
  ed.

\end{thebibliography}
\end{document}